\PassOptionsToPackage{prologue,table,xcdraw}{xcolor}
\documentclass[sigconf]{acmart}

\usepackage{multirow}
\usepackage[normalem]{ulem}
\useunder{\uline}{\ul}{}

\usepackage{enumitem}

\usepackage{amssymb}
\usepackage{pifont}
\usepackage{mathtools}
\usepackage{url}
\usepackage{multirow}
\usepackage{subfigure}
\usepackage{bm}
\usepackage{verbatim}
\usepackage{ragged2e}
\usepackage{caption}
\usepackage{subcaption}
\usepackage{graphicx}
\usepackage[normalem]{ulem}
\useunder{\uline}{\ul}{}
\usepackage{amsmath}
\usepackage{longtable}
\usepackage{adjustbox}
\usepackage{microtype}
\usepackage{setspace}
\usepackage{hyperref}
\usepackage{float}
\usepackage{balance}
\usepackage{algorithm}
\usepackage{algpseudocode}

\newcommand{\eg}{\emph{e.g., }}

\newcommand{\aka}

\AtBeginDocument{%
  \providecommand\BibTeX{{%
    \normalfont B\kern-0.5em{\scshape i\kern-0.25em b}\kern-0.8em\TeX}}}

\copyrightyear{2026}
\acmYear{2026}
\setcopyright{cc}
\setcctype{by}
\acmConference[KDD '26]{Proceedings of the 32nd ACM SIGKDD Conference on Knowledge Discovery and Data Mining V.2}{August 09--13, 2026}{Jeju Island, Republic of Korea}
\acmBooktitle{Proceedings of the 32nd ACM SIGKDD Conference on Knowledge Discovery and Data Mining V.2 (KDD '26), August 09--13, 2026, Jeju Island, Republic of Korea}
\acmDOI{10.1145/3770855.3817855}
\acmISBN{979-8-4007-2259-2/2026/08}

\begin{document}

\newcommand{\xyz}[1]{\textcolor{red}{#1}} 
\title{Intuition-Guided Latent Reasoning for LLM-Based Recommendation}


\author{Chang Liu}
\orcid{0009-0000-8324-7153}
\affiliation{%
  \institution{State Key Laboratory of Complex \& Critical Software Environment, Beihang University}
  \city{Beijing}
  \country{China}
}
\email{chang.liu@buaa.edu.cn}

\author{Yimeng Bai}
\orcid{0009-0008-8874-9409}
\affiliation{%
  \institution{University of Science and Technology of China}
  \city{Hefei}
  \country{China}
}
\email{baiyimeng@mail.ustc.edu.cn}

\author{Xiaoyan Zhao}
\orcid{0000-0001-6001-1260}
\affiliation{%
  \institution{The Chinese University of Hong Kong}
  \city{Hong Kong}
  \country{China}
}
\email{xzhao@se.cuhk.edu.hk}
\authornote{Corresponding author.}

\author{Yang Zhang}
\orcid{0000-0002-7863-5183}
\affiliation{%
 \institution{National University of Singapore}
 \city{Singapore}
 \country{Singapore}
}
\email{zhangy@nus.edu.sg}

\author{Qifan Wang}
\orcid{0000-0002-7570-5756}
\affiliation{
  \institution{Meta AI}
  \city{Menlo Park}
  \country{United States}
}
\email{wqfcr@fb.com}

\author{Fuli Feng}
\orcid{0000-0002-5828-9842}
\affiliation{
  \institution{University of Science and Technology of China}
  \city{Hefei}
  \country{China}
}
\email{fulifeng93@gmail.com}

\author{Wenge Rong}
\orcid{0000-0002-4229-7215}
\affiliation{%
  \institution{Beihang University}
  \city{Beijing}
  \country{China}
}
\email{w.rong@buaa.edu.cn}
\authornotemark[1]



\renewcommand{\shortauthors}{Chang Liu et al.}

\begin{abstract}

Large Language Models (LLMs) have demonstrated impressive reasoning capabilities in complex problem-solving tasks, motivating their use for preference reasoning in recommender systems. Latent reasoning, which operates in continuous hidden spaces rather than discrete tokens, has recently emerged as a promising paradigm for LLM-based recommendation. However, existing methods often start from unconstrained reasoning points, where hidden representations are misaligned with target item embeddings, leading to suboptimal reasoning trajectories.

Inspired by cognitive neuroscience, which suggests that human multi-step reasoning is guided by intuition as a latent prior, we propose \emph{IntuRec}, a two-stage framework that anchors latent reasoning with \emph{recommendation intuition}. In the extraction stage, the LLM-based recommender generates a top-$K$ candidate set based on users’ histories as the source of intuition. In the injection stage, the candidate set is transformed into a preference-aligned intuition embedding using self- and cross-attention mechanisms, which initializes the reasoning start point and guides subsequent latent reasoning. By providing a semantically grounded starting point, IntuRec efficiently explores the preference space along more accurate reasoning trajectories. Extensive experiments on multiple real-world datasets demonstrate that IntuRec consistently outperforms state-of-the-art baselines. We release our code at https://github.com/Ten-Mao/IntuRec.

\end{abstract}

\ccsdesc[500]{Information systems~Recommender systems}


\keywords{LLM-Based Recommendation; Latent Reasoning; Recommendation Intuition}


\maketitle
\newcommand\kddavailabilityurl{https://doi.org/10.5281/zenodo.20447119}
\ifdefempty{\kddavailabilityurl}{}{
\begingroup\small\noindent\raggedright\textbf{Resource Availability:}\\
The source code and artifacts of this paper have been made publicly available via Zenodo at \url{\kddavailabilityurl}, and the GitHub repository can be accessed at \url{https://github.com/Ten-Mao/IntuRec}.
\endgroup
}

\section{Introduction}

\begin{figure}[t]
    \centering
    \includegraphics[width=0.95\linewidth]{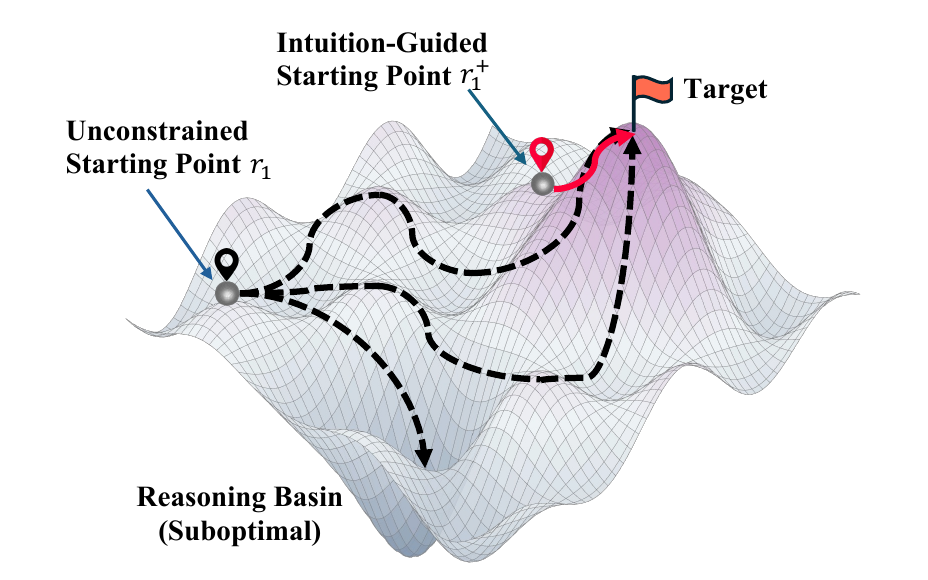}
    \caption{Latent reasoning start points of our proposed IntuRec versus baseline methods on the embedding space, showing how IntuRec initializes closer to target items. }
    \label{fig:r3_vs_tgt}
    \Description{..}
    \vspace{-5pt}
\end{figure}

Large Language Models (LLMs) have emerged as a transformative technology, redefining the artificial intelligence landscape by enabling in-context learning, instruction following, and multi-step reasoning at unprecedented scales~\cite{gpt4, gemini2.5, qwen3, seed1.5, openvla, deepseekv3, llama3}. Consequently, adapting LLMs for recommendation has become a prominent research frontier, garnering significant attention from both academia and industry~\cite{tallrec,lc-rec,hllm,openonerec,llm4rec_survey1,llm4rec_survey2,llm4rec_survey3,xiaoyan1,liu2025inference}. By leveraging their extensive world knowledge to enhance item representations and employing advanced reasoning to infer user preferences, LLM-based recommendation has the potential to emerge as a next-generation paradigm, fundamentally transforming the field of recommender systems~\cite{onerec-v2, onerec-v1, rec_instruct, llm2rec}.

Early studies in LLM-based recommendation primarily focused on fine-tuning LLMs to directly generate items from user history and context~\cite{bigrec,d3,llara,p5,p5-cid,igd}. However, such direct generation is often hindered by limited inferential depth and fails to fully exploit the reasoning potential of LLMs~\cite{deepseek-r1}. Recent studies have explored advanced reasoning techniques to address these limitations~\cite{recllm-r1,r2rec,onerec-think,reccot,cot4rec,reason4rec}. By adopting a ``think-before-recommend'' paradigm, these methods effectively decipher the latent rationale underlying evolving user interests and refine item-level semantic understanding, thereby yielding high-quality recommendations.

Among these approaches, \emph{latent reasoning} has gained prominence in LLM-based recommendation~\cite{rearec,lares,latentr3,plr,onepiece,s2gr}. Instead of explicitly decoding intermediate reasoning steps into discrete and human-interpretable tokens, these methods perform autoregressive inference directly within continuous latent spaces~\cite{reasoning_survey1,reasoning_survey2}, capturing complex user preferences with substantially fewer steps, thereby achieving significantly higher efficiency than CoT reasoning~\cite{simcot,coconut,codi}. The effectiveness of this paradigm is exemplified by OnePiece~\cite{onepiece}, which implements block-wise latent reasoning in Shopee’s ranking pipelines for personalized search, achieving consistent improvements across key business metrics.

Despite their efficiency, existing latent reasoning methods primarily supervise the final recommendation output, leaving the reasoning process itself largely unconstrained. This lack of guidance is most pronounced at the reasoning start point, which critically shapes the trajectory of multi-step latent reasoning~\cite{First-Step_Advantage}. As shown in Figure~\ref{fig:r3_vs_tgt}, reasoning start points are often poorly aligned with the target item embeddings, potentially biasing the reasoning path toward irrelevant or suboptimal regions and limiting recommendation performance. This motivates the need for explicit guidance at the reasoning start point to steer the latent reasoning process more reliably and effectively.

{To determine how such guidance should be structured, we draw on insights from cognitive neuroscience, which suggest that human multi-step reasoning is rarely initiated from an unconstrained deliberative process. Instead, it is preceded by \emph{intuition} — a latent, coarse-grained prior that narrows the solution space toward promising directions~\cite{goel2009cognitive,miller2001integrative,horr2014feeling,kok2013prior}. When solving complex problems, individuals first form a sense of plausible strategies and then reason within this restricted subspace. This intuition efficiently prunes the search space and mitigates early deviations.
In contrast, latent reasoning in current LLM-based recommendation models lacks such an internal prior. Token-level reasoning unfolds autoregressively without an explicit global constraint, making it prone to drifting toward suboptimal reasoning basins. Inspired by the human mechanism, we posit that recommendation reasoning should likewise be anchored by a \emph{recommendation intuition} to initialize the reasoning process, providing a directional embedding that modulates subsequent token-level reasoning trajectories toward the user's true underlying preferences.}

To this end, we propose \emph{IntuRec}, a two-stage framework that explicitly extracts and injects recommendation intuition for latent reasoning. In the extraction stage, the LLM-based recommender is trained under a sequential recommendation setting to predict the next item from users’ interaction histories. Beam search is then applied to generate an offline top-$K$ candidate list, which serves as the source of intuition for the next stage. In the injection stage, this candidate set---representing potential user preferences---is transformed into a single intuition embedding, which directly fills the reasoning start point. This intuition construction applies self-attention individually within each candidate item to capture its internal token-level structure, and cross-attention to integrate user and contextual information. By reasoning from this semantically grounded representation, IntuRec transcends blind exploration, efficiently and accurately capturing complex user preferences. Extensive experiments on multiple real-world datasets show that IntuRec consistently outperforms existing baselines.

The main contributions of this work are summarized as follows:
\begin{itemize}[leftmargin=*]
\item We introduce the concept of \emph{recommendation intuition} from a human reasoning perspective to guide latent reasoning in LLM-based recommendation.

\item We propose \emph{IntuRec}, a two-stage framework that extracts and injects recommendation intuition to initialize latent reasoning from a preference-aligned representation.

\item We conduct extensive experiments on multiple real-world datasets,
demonstrating the effectiveness of IntuRec.
\end{itemize}

\section{Preliminary}

In this section, we formally define the task of LLM-based recommendation and introduce latent reasoning, which underpins our method. We focus on \textit{sequential recommendation}, where the objective is to predict the next item a user is likely to interact with, based on their historical interactions~\cite{SASRec}.

\vspace{+3pt}
\textbf{LLM-Based Recommendation.} 
Formally, let $\mathcal{D}$ denote the recommendation dataset. Each instance is a triplet $(u, h, y) \in \mathcal{D}$, where $u$ is a user, $h$ represents the user’s historical interactions, and $y$ is the next item the user will interact with. Both $h$ and $y$ can be represented using textual information, such as item titles or descriptions~\cite{bigrec}. For each instance, the historical interactions $h$ are converted into a textual prompt $x$, which is then input to the LLM to generate the next-item recommendation:
\begin{equation}\label{eq:direct_gen}
    x \xrightarrow{\text{LLM}(x)} \hat{y},
\end{equation}
where $\hat{y}$ is the predicted next item. While simple, this direct generation approach often lacks reasoning depth and may fail to fully capture complex user preferences.

\vspace{+3pt}
\textbf{Latent Reasoning.} 
Instead of directly generating $\hat{y}$, latent reasoning introduces a sequence of intermediate states $\bm{r}=\{\bm{r}_n\}_{n=1}^N$ that progressively refine the model’s understanding of user preferences, where each $\bm{r}_n$ represents the $n$-th latent reasoning state embedding and $\bm{r}$ as a whole denotes the complete reasoning trajectory. The process can be written as:
\begin{equation}
    x \xrightarrow{\text{LLM}(x)} \bm{r} \xrightarrow{\text{LLM}(x, \bm{r})} \hat{y},
\end{equation}
with $N$ being the total number of reasoning steps. Each state is generated autoregressively by the LLM~\cite{latentr3}:
\begin{equation}\label{eq:latent_reasoning_original}
\begin{aligned}
\bm{r}_1 &= \text{LLM}(x)[-1], \\
\bm{r}_n &= \text{LLM}(x, \bm{r}_1, \dots, \bm{r}_{n-1})[-1], \quad n=2,\dots,N,
\end{aligned}
\end{equation}
where $[-1]$ denotes taking the last-position output. This shift from direct generation to latent reasoning reflects a transition from  quick inference (System-1 thinking) to deliberate reasoning (System-2 thinking) in LLM-based recommendation~\cite{kahneman2011thinking}.

\section{Methodology}
\begin{figure*}[t]
    \centering
    \includegraphics[width=0.98\textwidth]{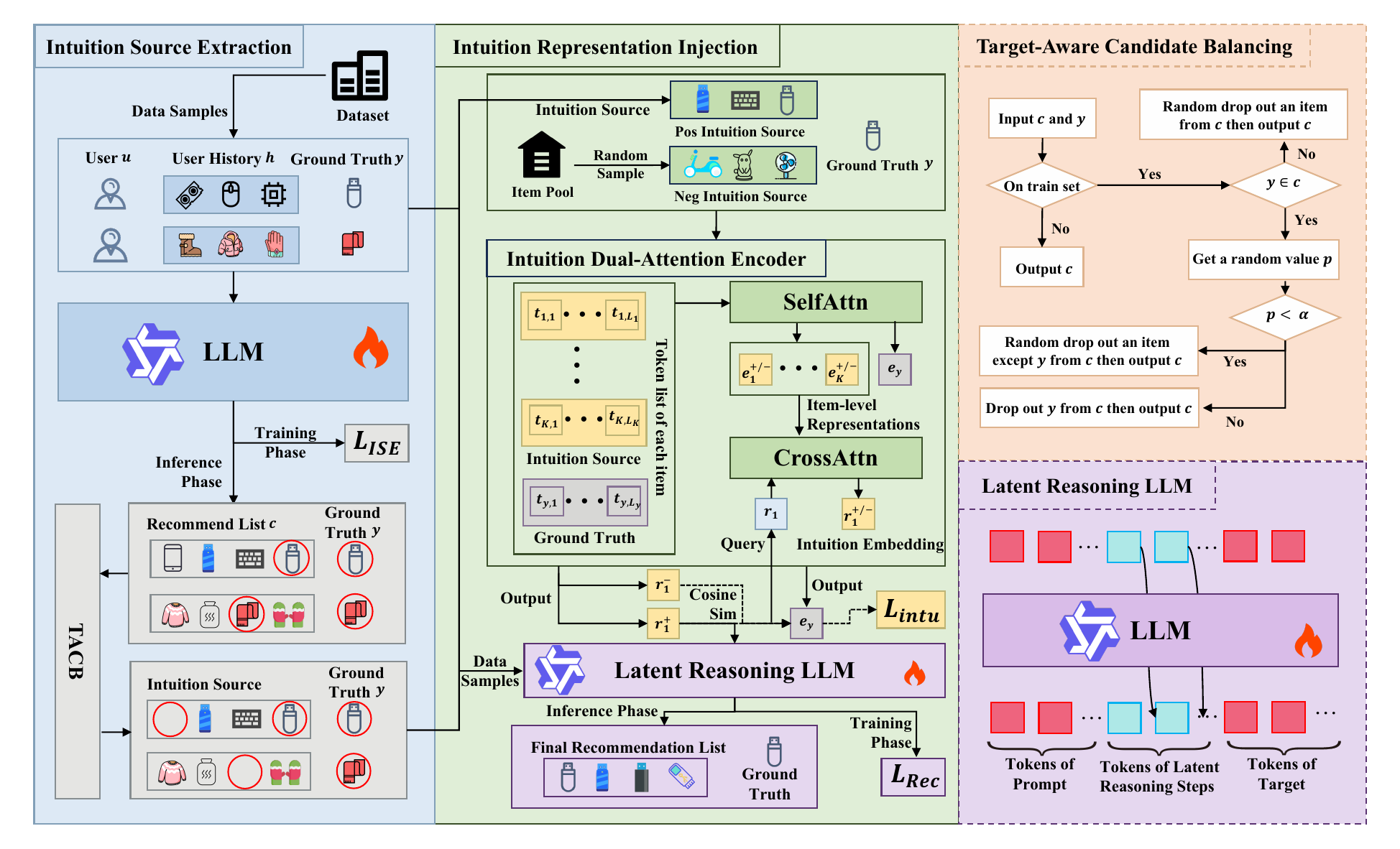}
    \caption{Overview of the proposed IntuRec framework. The two blocks on the left, ISE and IRI, illustrate the two-stage process, while the right side provides detailed internal views of the TACB and the latent reasoning LLM component. Specifically, IntuRec first executes ISE under a standard sequential recommendation setting to extract the intuition source. This source is then fed into IDAE to generate intuition embeddings optimized with the BPR loss. Finally, the LLM-based recommender performs latent reasoning under the guidance of these embeddings to produce the final results.}
    \label{fig:main}
    \Description{..}
\end{figure*}
In this section, we first provide an overview of the proposed methodology, followed by a detailed introduction of its components.

\subsection{Overview}

We aim to introduce explicit guidance for the reasoning starting point to improve the reliability and effectiveness of latent reasoning in LLM-based recommendation. This transforms the latent reasoning process from unguided exploration into a structured progression grounded in a principled initialization. To achieve this, our key belief is that emulating human reasoning by constructing and leveraging \emph{recommendation intuition}---which captures preference-aligned item patterns---provides a reliable cognitive prior for reasoning.

Building on this idea, we propose IntuRec, a novel framework that explicitly extracts and injects recommendation intuition to guide latent reasoning, as shown in Figure~\ref{fig:main}. IntuRec consists of two stages: (1) 
\emph{Intuition Source Extraction (ISE)}, which constructs a top-$K$ candidate set as a structured source of recommendation intuition, forming the foundation for the subsequent stage; and (2) \emph{Intuition Representation Injection (IRI)}, which encodes the candidate set into a unified intuition embedding and injects it as the reasoning start point to guide latent reasoning.

\subsection{Intuition Source Extraction (ISE)}
To construct the candidate item set as the intuition source, we train an LLM-based recommender in a sequential recommendation setting to predict the next item from users’ interaction histories. Formally, this follows Equation~\eqref{eq:direct_gen} and is optimized with the standard next-token prediction loss~\cite{bigrec}:
\begin{equation}\label{eq:loss_f1}
    \mathcal{L}_\text{ISE}= - \sum_{i=1}^{|y|} \log P_{\theta}(y_i \mid x, y_{<i}). 
\end{equation}
Here, $y_{<i}$ denotes the token sequence of the target item $y$ preceding the $i$-th token $y_i$. 

\subsubsection{Beam Search Generation}
After training, for each instance $(u,h,y)\in\mathcal{D}$, we perform beam search to generate a top-$K$ candidate list $c^+ = \{c_k^+\}_{k=1}^{K}$, which characterizes the patterns of items most likely to match user preferences. For comparison, we also construct a candidate list $c^-=\{c_k^-\}_{k=1}^{K}$ by randomly sampling items to represent less-preferred alternatives. The candidates are stored offline, expanding each original instance into $(u,h,c^+,c^-,y)$ for use in the subsequent injection stage. 

While candidate lists could in principle be obtained from other models such as SASRec~\cite{SASRec}, we generate them using the LLM-based recommender to maintain alignment with the downstream reasoning architecture, ensuring that the extracted recommendation intuition is fully compatible with the latent reasoning process and represents the model’s self-intuition.

\subsubsection{Target-Aware Candidate Balancing (TACB)}\label{sec:TACB}
A practical challenge in candidate construction is that the top-$K$ candidate list $c^+$ may contain the ground-truth target item $y$. An excessively high inclusion ratio can induce \emph{shortcut learning}~\cite{geirhos2020shortcut}, where the model over-relies on the direct presence of the target item instead of learning meaningful preference patterns, while an overly low ratio may weaken the supervision signal and harm training stability. 

To address this, we propose a Target-Aware Candidate Balancing (TACB) strategy that explicitly regulates the proportion of training instances whose candidate sets contain the ground-truth target item. For each training instance, we first perform beam search to generate a top-$(K\!+\!1)$ candidate list, and then construct the final top-$K$ candidate set as follows:
\begin{itemize}[leftmargin=*]
    \item If the candidate list does not contain the target item $y$, we randomly remove one candidate to form a size-$K$ set.
    \item If the candidate list contains $y$, we retain $y$ with probability $\alpha$ and randomly remove one of the remaining candidates; with probability $1-\alpha$, we remove $y$ and randomly discard one other candidate to form the size-$K$ set.
\end{itemize}

In our implementation, the retention probability $\alpha$ is explicitly calibrated to preserve distributional consistency between training and evaluation. Let $\beta_{\text{train}}$ denote the original proportion of training instances whose top-$(K\!+\!1)$ candidate lists naturally contain the ground-truth target item, and let $\beta_{\text{valid}}$ denote the corresponding proportion of validation instances whose top-$K$ candidate lists contain the target item. Formally, we set:
\begin{equation}\label{eq:balancing}
\alpha = \frac{\beta_{\text{valid}}}{\beta_{\text{train}}},
\end{equation}
so that, in expectation, the adjusted training distribution matches the target-item inclusion ratio observed during validation. This calibration avoids artificial distribution shifts introduced by candidate manipulation and ensures that training and evaluation are conducted under consistent candidate exposure conditions.

\subsection{Intuition Representation Injection (IRI)}

Once the intuition source is obtained, we integrate it into the latent reasoning process. Specifically, the top-$K$ candidate set $c^+$ is encoded into a compact embedding using a intuition encoder model $f(\cdot)$, which is then injected into the reasoning start point. Although the most straightforward approach might be to treat the top-$K$ candidate set $c^+$ as auxiliary text and append it directly to the prompt, we find that this often produces suboptimal results. This is likely because it is homogeneous with the original historical sequence, making its influence on the reasoning process too indirect. Formally, our revised latent reasoning process can be expressed as:
\begin{equation}\label{eq:latent_reasoning_new}
\begin{aligned}
\bm{r}_1 &= \text{LLM}(x)[-1], \\
\bm{r}^+_1 &= f(\bm{r}_1, c^+), \\
\bm{r}_n &= \text{LLM}(x, \bm{r}^+_1, \dots, \bm{r}_{n-1})[-1], \quad n=2,\dots,N.
\end{aligned}
\end{equation}
Compared with Equation~\eqref{eq:latent_reasoning_original}, this formulation replaces $\bm{r}_1^+$ with $\bm{r}^+_1$ and incorporates structured preference information into the initial latent state, providing semantically grounded explicit guidance that shapes the subsequent reasoning trajectory. In the following, we describe the design and implementation details of the intuition encoder model $f(\cdot)$ used to encode the candidate set.

\subsubsection{Intuition Dual-Attention Encoder (IDAE)}
The intuition encoder model $f(\cdot)$ transforms the top-$K$ candidate set into a structured intuition embedding that guides subsequent latent reasoning process. First, each candidate item is tokenized, and its corresponding LLM token embeddings are obtained. Let $\bm{t}_{k,l}$ denote the embedding of the $l$-th token in the $k$-th candidate $\bm{c}_k^+$. To capture the internal structure of each item, self-attention is applied individually to the token embeddings of each item, followed by last-token pooling operation to produce a compact item-level representation:
\begin{equation}\label{eq:self-attn}
\bm{e}_k^+ = \text{SelfAttn}({\{\bm{t}_{k,l}\}_{l=1}^{L_k}})[-1], \quad k=1,\dots,K,
\end{equation}
where $L_k$ denotes the number of tokens corresponding to the candidate item $\bm{c}_k^+$, and $\bm{e}_k^+$ represents the resulting item embedding after intra-item self-attention and pooling.

Next, these item-level representations are integrated with the initial reasoning state $\bm{r}_1$ via cross-attention, where $\bm{r}_1$ serves as the query and ${\bm{e}_1^+, \dots, \bm{e}_K^+}$ serve as keys and values:
\begin{equation}\label{eq:cross-attn}
\bm{r}_1^+ = \text{CrossAttn}(Q=\bm{r}_1, K=\{\bm{e}_k^+\}_{k=1}^{K}, V=\{\bm{e}_k^+\}_{k=1}^{K}).
\end{equation}
This ensures that the intuition embedding $\bm{r}_1^+$ is directly conditioned on the initial reasoning state, effectively injecting structured candidate information into the reasoning start point while maintaining alignment with user history and context. Finally, the enriched latent state $\bm{r}_1^+$ is used to replace the original reasoning start point $\bm{r}_1$ according to Equation~\eqref{eq:latent_reasoning_new}.

\subsubsection{Optimization Strategy}

The LLM recommender is trained to predict the next item in a user’s interaction sequence, guided by the enriched latent reasoning states. Formally, the primary training objective is the standard next-token prediction loss:  
\begin{equation}\label{eq:loss_rec}
    \mathcal{L}_\text{rec}= - \sum_{i=1}^{|y|} \log P_{\theta}(y_i \mid x, \bm{r}^+, y_{<i}),
\end{equation}
where $y_{<i}$ denotes the sequence of tokens preceding the $i$-th token $y_i$ of the target item $y$, $x$ represents the input prompt, and $\bm{r}^+ = \{\bm{r}_1^+, \bm{r}_2, \dots, \bm{r}_N\}$ is the enriched latent state incorporating recommendation intuition extracted by the IDAE module. This loss ensures that the LLM leverages the structured guidance provided by the candidate set when generating the next item.

\vspace{+3pt}
\textbf{Intuition Alignment Loss.}  
To further guide the IDAE module and strengthen semantic consistency between the candidate-guided intuition and the target item, we introduce a contrastive intuition alignment loss following the Bayesian Personalized Ranking (BPR) paradigm~\cite{BPR}. Specifically, the positive latent state $\bm{r}_1^+$ is encouraged to be more similar to the target item embedding than a negative counterpart $\bm{r}_1^-$, effectively providing a preference-aligned supervision signal.  

The embedding of the ground-truth target item is obtained using the same IDAE mechanism applied to candidate items, i.e., intra-item self-attention followed by last-token pooling:  
\begin{equation}
\bm{e}_y = \text{SelfAttn}(\{\bm{t}_{y,l}\}_{l=1}^{L_y})[-1],
\end{equation}
where $\bm{t}_{y,l}$ denotes the token embedding of the $l$-th token. Negative latent states $\bm{r}_1^-$ are constructed analogously from randomly sampled items $c^-$ using the same IDAE procedure. The contrastive intuition loss is defined as:  
\begin{equation}\label{eq:loss_intu}
\mathcal{L}_\text{intu} = -\log \sigma \Big( \text{sim}(\bm{r}_1^+, \bm{e}_y) - \text{sim}(\bm{r}_1^-, \bm{e}_y) \Big),
\end{equation}
where $\sigma(\cdot)$ is the sigmoid function and $\text{sim}(\cdot, \cdot)$ denotes cosine similarity. This additional supervision prevents the IDAE module from producing embeddings that are structurally plausible but semantically biased, which could result from training solely on high-quality top-$K$ candidates and may lead to suboptimal latent reasoning and recommendation performance.

\vspace{+3pt}
\textbf{Overall Loss}.
The IDAE module is trained jointly with the LLM recommender in an end-to-end manner, enabling unified optimization of recommendation intuition construction and latent reasoning. The overall training objective is defined as a weighted combination of the recommendation loss and the intuition alignment loss:
\begin{equation}\label{eq:overall_loss}
\mathcal{L}_\text{IRI} = \mathcal{L}_\text{rec} + \lambda \, \mathcal{L}_\text{intu},
\end{equation}
where $\lambda$ is a hyperparameter that balances next-item prediction and intuition alignment.

\subsection{Training Pipeline}
\begin{algorithm}[t]
\caption{Two-Stage Training of IntuRec}
\label{alg:training_inturec}
\begin{algorithmic}[1]
\While{not converged} \Comment{\textbf{Stage ISE}}
    \State Compute $\mathcal{L}_\text{ISE}$ (Equation~\eqref{eq:loss_f1});
    \State Update $\theta_\text{LLM}$;
\EndWhile
\State Generate $c^+, c^-$ with TACB;
\While{not converged} \Comment{\textbf{Stage IRI}}
    \State Encode $c^+$ via IDAE to get $\bm{r}_1^+$;
    \State Encode target item $y$ via IDAE to get $\bm{e}_y$;
    \State Encode negatives $c^-$ via IDAE to get $\bm{r}_1^-$;
    \State Compute $\mathcal{L}_\text{intu}$ (Equation~\eqref{eq:loss_intu});
    \State Inject $\bm{r}_1^+$ into LLM latent reasoning (Equation~\eqref{eq:latent_reasoning_new});
    \State Compute $\mathcal{L}_\text{rec}$ (Equation~\eqref{eq:loss_rec});
    \State Compute $\mathcal{L}_\text{IRI}$ (Equation~\eqref{eq:overall_loss});
    \State Update $\theta_\text{LLM}$ and $\theta_\text{IDAE}$;
\EndWhile
\end{algorithmic}
\end{algorithm}

We summarize the overall training procedure of IntuRec in Algorithm~\ref{alg:training_inturec}, which proceeds in two progressive stages:

In Stage ISE (lines 1–5), the LLM-based recommender is trained to predict the next item from user interaction histories. At each iteration, the next-token prediction loss $\mathcal{L}_\text{ISE}$ (Equation~\eqref{eq:loss_f1}) is computed (line 2) and used to update the parameters $\theta_\text{LLM}$ (line 3) until convergence. Once training is completed, the top-$K$ positive and random negative candidate sets $c^+$ and $c^-$ are generated with the TACB strategy in Section~\ref{sec:TACB} (line 5).

In Stage IRI (lines 6–15), the framework focuses on learning the candidate-guided intuition embedding and integrating it into the latent reasoning process. First, the positive candidate set $c^+$ is encoded via the IDAE module to produce the enriched latent state $\bm{r}_1^+$ (line 7). The ground-truth target item $y$ is also encoded via IDAE to obtain $\bm{e}_y$ (line 8), and the negative candidate set $c^-$ is encoded to obtain $\bm{r}1^-$ (line 9). Using these representations, the contrastive intuition alignment loss $\mathcal{L}_\text{intu}$ (Equation~\eqref{eq:loss_intu}) is computed (line 10). The enriched intuition embedding $\bm{r}_1^+$ is then injected into the LLM’s latent reasoning according to Equation~\eqref{eq:latent_reasoning_new} (line 11), and the recommendation loss $\mathcal{L}_\text{rec}$ (Equation~\eqref{eq:loss_rec}) is calculated (line 12). Subsequently, the overall training objective $\mathcal{L}_\text{IRI}$ (Equation~\eqref{eq:overall_loss}) is computed (line 13), and the parameters of both the LLM $\theta_\text{LLM}$ and the IDAE module $\theta_\text{IDAE}$ are updated using gradient-based optimization (line 14). This procedure continues iteratively until convergence, enabling joint optimization of recommendation prediction and intuition-guided latent reasoning.

\section{Experiment}

In this section, we conduct a series of experiments to answer the following research questions:\\
\textbf{RQ1:} How does the proposed IntuRec framework perform compared to existing latent reasoning recommendation methods? \\
\textbf{RQ2:} What is the contribution of each component within IntuRec to the overall performance? \\
\textbf{RQ3:} Does IntuRec extract intuition embeddings that serve as a more advantageous initialization compared to other models? \\
\textbf{RQ4:} How do specific hyper-parameters influence IntuRec?

\subsection{Experimental Setting}
\label{exp_settings}
\subsubsection{Datasets}
\label{exp_datasets}

\begin{table}[t]
\centering
\caption{ Statistical details of the evaluation datasets.}
\label{tab:data}
\resizebox{\columnwidth}{!}{%
\begin{tabular}{ccccccc}
\hline
Dataset & \#Train & \#Valid & \#Test & \#User num & \#Item num & Sparsity \\ \hline
CDs     & 49251 & 6156  & 6158 & 7685     & 5841     & 99.86\%  \\
Toys    & 53898 & 6737  & 6738 & 11311    & 6299     & 99.91\%  \\
Games   & 75175 & 9397  & 9397 & 13577    & 5308     & 99.87\%  \\ \hline
\end{tabular}%
}
\end{table}

\begin{table*}[t]
\centering
\caption{Overall performance comparison between baselines and our proposed IntuRec. Specifically, IntuRec is applied on top
of BIGRec and D$^3$–referred to as IntuRec-B and IntuRec-D and LatentR$^3$ is applied on top of BIGRec and D$^3$–referred to as LatentR$^3$-B and LatentR$^3$-D. The best and second-best results are highlighted in bold and underlined, respectively.}
\label{tab:main_tab}
\resizebox{\textwidth}{!}{%
\begin{tabular}{clcccccccccccccc}
\hline
 &
  \multicolumn{1}{c}{} &
  \multicolumn{4}{c}{Traditional Methods} &
   &
  \multicolumn{9}{c}{LLM-based Methods} \\ \cline{3-6} \cline{8-16} 
\multirow{-2}{*}{Dataset} &
  \multicolumn{1}{c}{\multirow{-2}{*}{Metric}} &
  Caser &
  GRU4Rec &
  SASRec &
  ReaRec &
   &
  LLM-Base &
  LLM-CoT &
  AlphaRec &
  BIGRec &
  LatentR$^3$-B &
  \cellcolor[HTML]{EFEFEF}IntuRec-B &
  D$^3$ &
  LatentR$^3$-D &
  \cellcolor[HTML]{EFEFEF}IntuRec-D \\ \hline
 &
  Recall@5 &
  0.0469 &
  0.0481 &
  0.0841 &
  0.0845 &
   &
  0.0195 &
  0.0302 &
  0.0479 &
  0.0757 &
  0.0918 &
  \cellcolor[HTML]{EFEFEF}0.0986 &
  {\ul 0.1122} &
  0.1077 &
  \cellcolor[HTML]{EFEFEF}\textbf{0.1174} \\
 &
  Recall@10 &
  0.0689 &
  0.0669 &
  0.1054 &
  0.01057 &
   &
  0.0252 &
  0.0406 &
  0.0774 &
  0.0929 &
  0.1139 &
  \cellcolor[HTML]{EFEFEF}0.1221 &
  0.1272 &
  {\ul 0.1306} &
  \cellcolor[HTML]{EFEFEF}\textbf{0.1368} \\
 &
  NDCG@5 &
  0.0312 &
  0.0365 &
  0.0622 &
  0.0670 &
   &
  0.0148 &
  0.0213 &
  0.0278 &
  0.0616 &
  0.0756 &
  \cellcolor[HTML]{EFEFEF}0.0795 &
  {\ul 0.0906} &
  0.0886 &
  \cellcolor[HTML]{EFEFEF}\textbf{0.0933} \\
\multirow{-4}{*}{CDs} &
  NDCG@10 &
  0.0382 &
  0.0425 &
  0.0691 &
  0.0737 &
   &
  0.0167 &
  0.0246 &
  0.0373 &
  0.0672 &
  0.0828 &
  \cellcolor[HTML]{EFEFEF}0.0871 &
  0.0955 &
  {\ul 0.0960} &
  \cellcolor[HTML]{EFEFEF}\textbf{0.0995} \\ \hline
 &
  Recall@5 &
  0.0251 &
  0.0417 &
  0.0601 &
  0.0565 &
   &
  0.0203 &
  0.0261 &
  0.0579 &
  0.0701 &
  0.0773 &
  \cellcolor[HTML]{EFEFEF}0.0858 &
  0.0830 &
  {\ul 0.0880} &
  \cellcolor[HTML]{EFEFEF}\textbf{0.0922} \\
 &
  Recall@10 &
  0.0384 &
  0.0564 &
  0.0760 &
  0.0757 &
   &
  0.0359 &
  0.0496 &
  0.0893 &
  0.0931 &
  0.1060 &
  \cellcolor[HTML]{EFEFEF}0.1077 &
  0.1026 &
  \textbf{0.1140} &
  \cellcolor[HTML]{EFEFEF}{\ul 0.1122} \\
 &
  NDCG@5 &
  0.0170 &
  0.0305 &
  0.0458 &
  0.0410 &
   &
  0.0128 &
  0.0153 &
  0.0347 &
  0.0508 &
  0.0573 &
  \cellcolor[HTML]{EFEFEF}{\ul 0.0662} &
  0.0610 &
  0.0655 &
  \cellcolor[HTML]{EFEFEF}\textbf{0.0708} \\
\multirow{-4}{*}{Toys} &
  NDCG@10 &
  0.0214 &
  0.0352 &
  0.0510 &
  0.0472 &
   &
  0.0178 &
  0.0229 &
  0.0448 &
  0.0582 &
  0.0665 &
  \cellcolor[HTML]{EFEFEF}0.0733 &
  0.0674 &
  {\ul 0.0740} &
  \cellcolor[HTML]{EFEFEF}\textbf{0.0773} \\ \hline
 &
  Recall@5 &
  0.0324 &
  0.0322 &
  0.0416 &
  0.0576 &
   &
  0.0236 &
  0.0120 &
  0.0558 &
  0.0461 &
  0.0590 &
  \cellcolor[HTML]{EFEFEF}0.0687 &
  0.0608 &
  {\ul 0.0713} &
  \cellcolor[HTML]{EFEFEF}\textbf{0.0751} \\
 &
  Recall@10 &
  0.0538 &
  0.0517 &
  0.0633 &
  0.0856 &
   &
  0.0311 &
  0.0194 &
  0.0893 &
  0.0709 &
  0.0884 &
  \cellcolor[HTML]{EFEFEF}0.0961 &
  0.0860 &
  {\ul 0.0977} &
  \cellcolor[HTML]{EFEFEF}\textbf{0.1033} \\
 &
  NDCG@5 &
  0.0211 &
  0.0207 &
  0.0280 &
  0.0397 &
   &
  0.0190 &
  0.0082 &
  0.0397 &
  0.0334 &
  0.0419 &
  \cellcolor[HTML]{EFEFEF}0.0488 &
  0.0423 &
  {\ul 0.0495} &
  \cellcolor[HTML]{EFEFEF}\textbf{0.0533} \\
\multirow{-4}{*}{Games} &
  NDCG@10 &
  0.0280 &
  0.0270 &
  0.0350 &
  0.0487 &
   &
  0.0214 &
  0.0105 &
  0.0515 &
  0.0414 &
  0.0513 &
  \cellcolor[HTML]{EFEFEF}0.0576 &
  0.0505 &
  {\ul 0.0580} &
  \cellcolor[HTML]{EFEFEF}\textbf{0.0625} \\ \hline
\end{tabular}%
}
\end{table*}

We conduct experiments on three subsets of the Amazon Review dataset\footnote{\url{https://nijianmo.github.io/amazon/index.html}}~\cite{Amazon18}: CDs, Toys and Games. The detailed statistics of these datasets are presented in Table~\ref{tab:data}. 

Following prior work, we first apply 5-core filtering, which removes users and items with fewer than five interactions. We then adopt the same dynamic temporal partitioning strategy as in~\cite{latentr3}. Specifically, the dataset is processed using a sliding time window with a fixed end time of October 2018, initially starting from October 2017. After 5-core filtering, if the resulting dataset contains fewer than 5,000 items, the start time is shifted backward by three months (\eg to July 2017). This procedure is repeated until the processed dataset contains more than 5,000 items, while keeping the end time fixed at October 2018. The resulting dataset is split into training, validation, and test sets with a ratio of 8:1:1. For model training, each user’s interaction sequence is truncated or padded to a fixed length of 10, retaining the most recent interactions.

\subsubsection{Baselines}
\label{exp_baselines}
The baselines for comparison fall into two categories: traditional and LLM-based recommendation approaches\footnote{Notably, both ReaRec and LatentR$^3$ implement \emph{latent reasoning} in their architectures.}. 

\vspace{+3pt}
(a) \emph{Traditional Recommendation Methods}:
\begin{itemize}[leftmargin=*]
    \item \textbf{Caser}~\cite{Caser} employs convolutional neural networks (CNNs) to extract sequential patterns from user behavior data.
    \vspace{+3pt}
    \item \textbf{GRU4Rec}~\cite{gru4rec} employs gated recurrent units (GRUs) for session-based recommendation.
    \vspace{+3pt}
    \item \textbf{SASRec}~\cite{SASRec} employs self-attention mechanisms within a decoder-only Transformer architecture to model user preferences.
    \vspace{+3pt}
    \item \textbf{ReaRec}~\cite{rearec} extends SASRec with multi-step autoregressive \emph{latent reasoning}; we adopt the Ensemble Reasoning Learning (ERL) variant for its more stable performance.
\end{itemize}

\vspace{+3pt}
(b) \emph{LLM-based Recommendation Methods}:
\begin{itemize}[leftmargin=*]
    \item \textbf{LLM-Base} \cite{qwen2.5} directly applies a pre-trained LLM to the recommendation task by feeding in the user’s interaction history and a prompt template, without any task-specific tuning. 
    \vspace{+3pt}
    \item \textbf{LLM-CoT} \cite{Rec-SAVER} extends LLM-Base by leveraging a zero-shot chain-of-thought (CoT) reasoning prompt to enhance recommendation performance.
    \vspace{+3pt}
    \item \textbf{AlphaRec} \cite{AlphaRec} builds a simple recommender model based on embeddings from LLMs.
    \vspace{+3pt}
    \item \textbf{BIGRec} \cite{bigrec} fine-tunes an LLM for recommendation, decoding via dot-product with vectorized item embeddings.
    \vspace{+3pt}
    \item  \textbf{D$^3$} \cite{d3} extends BIGRec by disabling length normalization for ghost tokens to reduce amplification bias and leveraging a text-free assistant to promote less frequent tokens, mitigating recommendation homogeneity.
    \vspace{+3pt}
    \item \textbf{LatentR$^3$} \cite{latentr3} introduces multi-step \emph{latent reasoning}, using an additional self-attention layer to encode reasoning tokens, which can be applied to both BIGRec and D$^3$, denoted as LatentR$^3$-B and LatentR$^3$-D, respectively.
\end{itemize}

\subsubsection{Evaluation Metrics}
\label{exp_metrics}
To ensure a fair comparison, we adopt two widely used evaluation metrics: Recall@$K$ and NDCG@$K$ with $K=5,10$. For LLM-based models, we conduct beam search with a beam size of 10 and truncate the resulting recommendation list to the desired $K$. To mitigate hallucinations during decoding, we further apply constrained decoding techniques~\cite{p5-cid}.

\subsubsection{Implementation Details}
For traditional ID-based baselines, we optimize using Adam, with learning rates tuned from \{1$e$-2, 1$e$-3, 1$e$-4\} and weight decay adjusted from \{1$e$-4, 1$e$-5, 1$e$-6\}, employing Binary Cross-Entropy loss with randomly sampled negative items. For LLM-based baselines, we use Qwen2.5‑1.5B \cite{qwen2.5} as the backbone, trained with AdamW, with learning rates selected from \{3$e$-3, 3$e$-4, 3$e$-5\}, and apply early stopping with a patience of 1. All other hyperparameters follow the official baseline implementations. For IntuRec, the candidate list size \(K\) is adjusted over \(\{5, 10, 15, 20, 25, 30\}\), the number of reasoning steps \(N\) is tuned over \(\{1, 2, 3, 4, 5\}\), and the loss balancing coefficient \(\lambda\) is varied over \(\{0.5, 1.0\}\). All experiments are conducted on RTX 4090 GPUs, except for hyperparameter studies with reasoning steps $N \ge 2$ (Section~\ref{exp_hyper}), which are run on A100 GPUs due to memory constraints.

\subsection{Performance Comparison (RQ1)}
\label{exp_overall}

We begin by assessing the overall recommendation performance of the compared methods on all three datasets. The summarized results are presented in Table~\ref{tab:main_tab}, yielding the following key observations:

\begin{itemize}[leftmargin=*]
\item Integrating IntuRec consistently improves performance across all backbones and baselines, with IntuRec-D achieving state-of-the-art results. These improvements stem from IntuRec’s ability to extract and inject recommendation intuition, initializing the latent reasoning process from a preference-aligned representation and effectively grounding the reasoning start point.

\item Overall, LatentR$^3$-B outperforms BIGRec, and LatentR$^3$-D outperforms D$^3$, demonstrating that latent reasoning can effectively increase the model’s computational depth, thereby enabling more comprehensive exploration of the item space and improving recommendation accuracy.

\item LLM-based recommendation methods generally outperform traditional ID-based approaches. This advantage stems from the LLMs’ ability to leverage world knowledge beyond what is captured in sparse user–item interactions, providing richer contextual information and enabling more informed decisions.
\end{itemize}

\subsection{Ablation Study (RQ2)}
\label{exp_ablation}

To validate the design rationale behind IntuRec, we perform a thorough evaluation by systematically ablation of each key component, yielding a set of model variants. Specifically, starting from IntuRec-B, we introduce the following variants for comparison:

\vspace{+3pt}

For the \textbf{Intuition Source Extraction (ISE)} stage:
\begin{itemize}[leftmargin=*]
    \item \textbf{w/o TACB}: This variant removes the Target-Aware Candidate Balancing strategy and directly generate top-$K$ candidate lists without performing the balancing operation.
    \item \textbf{w/ random intu}: This variant replaces the positive intuition source $c^+$ with the negative source $c^-$ in Equation~\eqref{eq:latent_reasoning_new}.
    \item \textbf{w/ low-pop intu}: This variant replaces the positive intuition source $c^+$ in Equation~\eqref{eq:latent_reasoning_new} with a list of items randomly sampled from the bottom 5\% least popular items in the candidate set.
\end{itemize}

For the \textbf{Intuition Representation Injection (IRI)} stage:
\begin{itemize}[leftmargin=*]
    \item \textbf{w/o IRI}: This variant completely removes the stage-two model training, relying solely on the LLM trained in ISE stage.
    \item \textbf{w/ text injection}: This variant directly embeds $c^+$ into the prompt in textual form, instead of injecting it as an embedding.
    \item \textbf{w/o SelfAttn}: This variant removes the self-attention layer in IDAE and directly applies mean pooling in Equation~\eqref{eq:self-attn}.
    \item \textbf{w/o CrossAttn}: This variant removes the cross-attention layer in IDAE and directly performs mean pooling in Equation~\eqref{eq:cross-attn}, which are then added to the query.
    \item \textbf{w/o $\mathcal{L}_\text{intu}$}: This variant removes the BPR loss that supervises the intuition embeddings in Equation~\eqref{eq:overall_loss}, relying solely on the recommendation loss for end-to-end training.

\end{itemize}

\begin{table}[t]
\centering
\caption{Ablation study of IntuRec-B on CDs dataset.}
\label{tab:ablation}
\resizebox{\columnwidth}{!}{%
\begin{tabular}{
>{\columncolor[HTML]{FFFFFF}}c 
>{\columncolor[HTML]{FFFFFF}}c 
>{\columncolor[HTML]{FFFFFF}}c 
>{\columncolor[HTML]{FFFFFF}}c 
>{\columncolor[HTML]{FFFFFF}}c }
\hline
Method                             & Recall@5        & Recall@10       & NDCG@5          & NDCG@10         \\ \hline
IntuRec-B                          & \textbf{0.0986} & \textbf{0.1221} & \textbf{0.0795} & \textbf{0.0871} \\ \hline
{-w/o TACB}               & 0.0949          & 0.1187          & 0.0777          & 0.0854          \\
{-w/ random intu}  & {\color[HTML]{1F2329} 0.0953} & {\color[HTML]{1F2329} 0.1156} & {\color[HTML]{1F2329} 0.0771} & {\color[HTML]{1F2329} 0.0835} \\
{-w/ low-pop intu} & {\color[HTML]{1F2329} 0.0937} & {\color[HTML]{1F2329} 0.1143} & {\color[HTML]{1F2329} 0.0776} & {\color[HTML]{1F2329} 0.0841} \\ \hline
{-w/o IRI}                & 0.0822          & 0.1015          & 0.0669          & 0.0731          \\ 
{-w/ text injection}       & 0.0762          & 0.0962          & 0.0601          & 0.0667          \\
{-w/o SelfAttn}           & 0.0963          & 0.1189          & 0.0776          & 0.0849          \\
{-w/o CrossAttn}          & 0.0892          & 0.1098          & 0.0718          & 0.0784          \\
{-w/o $\mathcal{L}_\text{intu}$} & 0.0957          & 0.1194          & 0.0787          & 0.0863          \\ \hline
\end{tabular}%
}
\end{table}

Table~\ref{tab:ablation} illustrates the comparison results on CDs, from which we draw the following observations:
\begin{itemize}[leftmargin=*]
    \item \textbf{ISE stage:} Removing TACB, or replacing the positive intuition source $c^+$ with random or low-popularity items, leads to notable performance drops. This indicates that constructing the recommendation intuition source using top-$K$ candidates is effective and provides high-quality starting points that better guide the latent reasoning process.
    \item \textbf{IRI stage:} (1) Completely removing IRI (- w/o IRI) demonstrates the necessity of latent reasoning for performance gains. (2) Embedding $c^+$ as plain text (- w/ text injection) shows that injecting intuition embeddings is more effective than a text-based approach that is homogeneous with the user’s historical interactions, providing more direct guidance for reasoning. (3) Removing SelfAttn or CrossAttn within the IDAE module leads to significant performance drops, indicating that these attention mechanisms are critical for processing information from the intuition source. (4) Removing the BPR loss supervising intuition embeddings (- w/o $\mathcal{L}_\text{intu}$) results in a noticeable performance decline, confirming that the introduced contrastive supervision is essential to ensure that the learned intuition embeddings remain aligned with user preferences.
\end{itemize}

\subsection{Visualization Analysis (RQ3)}
\label{exp_visual}

\begin{figure}[t]
    \centering
    \includegraphics[width=0.8\linewidth]{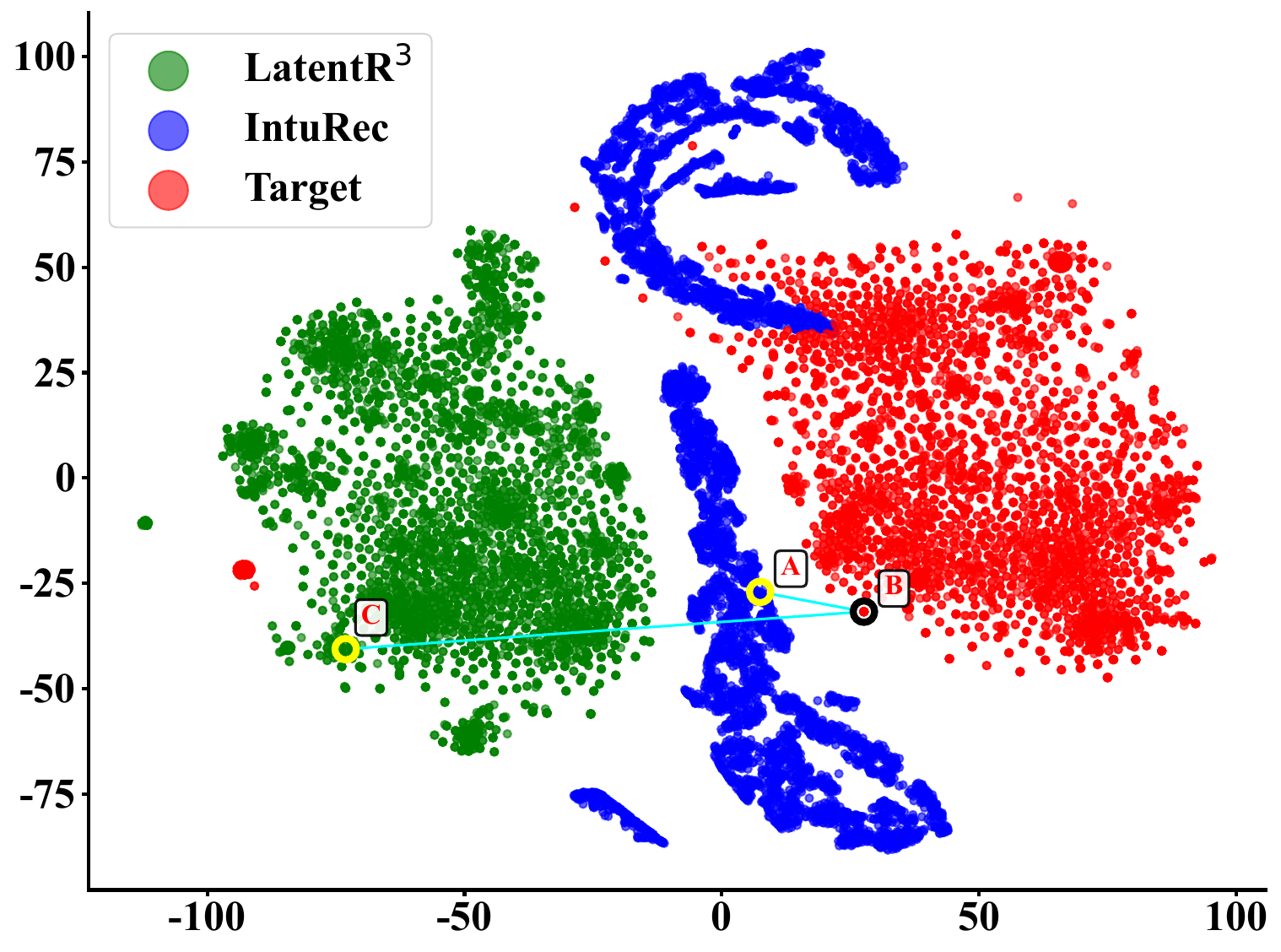}
    \caption{Embedding space distribution of reasoning start points of IntuRec (semantically guided by intuition embeddings) and LatentR$^3$ (unconstrained), along with target items, under cosine distance on the CDs dataset.}
    \label{fig:intu_vs_others}
    \Description{..}
\end{figure}

To gain deeper insights into the effectiveness of IntuRec, we conduct a visualization analysis of the embedding space distributions of the IntuRec intuition embedding $\bm{r}_1^+$, LatentR$^{3}$ reasoning start points $\bm{r}_1$, and target item embeddings $\bm{e}_y$ under cosine distance on the CDs dataset. The visualization results are shown in Figure~\ref{fig:intu_vs_others}, from which several key observations can be made:

\begin{itemize}[leftmargin=*]
    \item The embedding space of reasoning start points shows that IntuRec (semantically guided by intuition embeddings) lies substantially closer to the ground-truth target items compared to unconstrained LatentR$^3$ start points. For a representative example, point $A$ (IntuRec) is markedly nearer to the ground-truth point $B$ than point $C$ (LatentR$^3$, unconstrained). This evidence demonstrates that intuition embeddings provide a superior initialization for latent reasoning, effectively bridging the semantic gap from the outset and anchoring the reasoning process within a relevant preference manifold.

    \item More interestingly, the distribution of intuition embeddings is both target-oriented and highly anisotropic, indicating that the model has learned a compact preference subspace. Prior work~\cite{presentation_theory} has shown that when representation learning is driven by cosine-based contrastive objectives—particularly under strong feature selection bias—embeddings tend to form local manifolds rather than being uniformly distributed across the hypersphere. Similarly, in our method, the BPR loss used in Equation~\eqref{eq:loss_intu} encourages embeddings to align with directions corresponding to users’ historical behaviors. This provides an intuitive explanation for why our intuition embeddings are largely confined to specific subspaces, often forming curvilinear or ribbon-like manifolds. Consequently, we can view the recommendation intuition as a \textbf{``compass''}, effectively guiding latent reasoning along the learned manifolds and thereby substantially improving both search efficiency and recommendation accuracy.
\end{itemize}

\subsection{Hyperparameter Analysis (RQ4)}
\label{exp_hyper}

In our investigation, the two hyper-parameters $K$ and $N$ assume
pivotal roles in influencing the performance. Specifically, $K$ controls the number of candidate items in the intuition source, while $N$ determines the number of steps for implicit reasoning. We conducted systematic experiments on these two hyperparameters to assess their impact on model performance. We selected $K$ from the range $\{5, 10, 15, 20, 25, 30\}$ and $N$ from $\{1, 2, 3, 4, 5\}$. 

\begin{figure}[t]
    \centering
    \includegraphics[width=\linewidth]{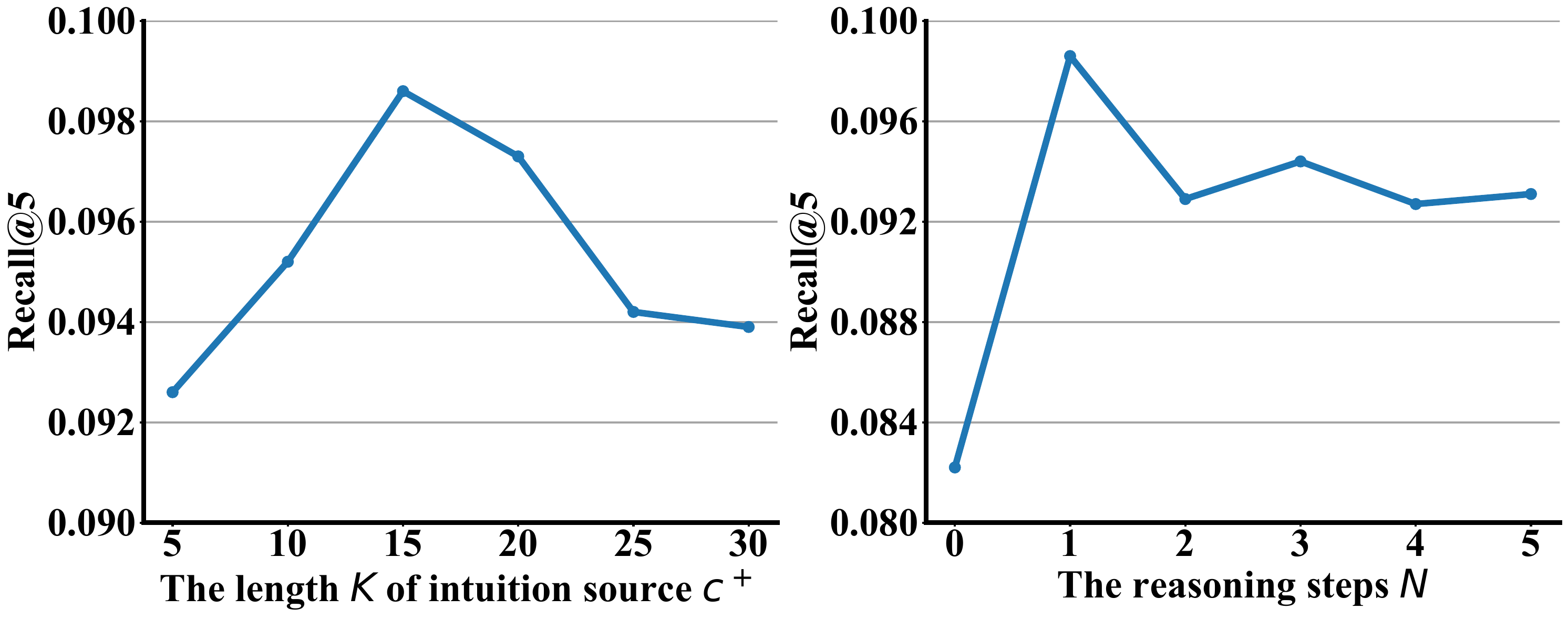}
    \caption{Hyperparameter study of IntuRec: the left plot shows the model's Recall@5 with respect to the hyperparameter $K$, and the right plot shows the model's Recall@5 with respect to the hyperparameter $N$.}
    \label{fig:hyper}
    \Description{..}
\end{figure}

The results are summarized in Figure~\ref{fig:hyper}, from which we can draw the following observations:
\begin{itemize}[leftmargin=*]
    \item As $K$ increases, the model's performance follows a monotonic upward trend, peaking at $K=15$ before monotonically declining. This behavior occurs because, at lower values of $K$, increasing the candidate count enhances diversity, allowing the intuition embedding to progressively approximate the ground-truth distribution. However, as $K$ becomes excessively large, the inherent decoding bias of LLM-based models takes over~\cite{d3}; the diversity of the intuition source plateaus, and the model may begin to repetitively emphasize specific items. This induces a human-like cognitive bias that hampers the model's exploratory capacity.
    \item As $N$ increases, the model reaches its peak performance immediately at $N=1$ and then exhibits a slight decline. This indicates that, given a high-quality starting point for reasoning, only a minimal number of reasoning steps is required to achieve optimal performance. In contrast, increasing the number of reasoning steps beyond this point can introduce additional challenges, as the model must maintain consistency across more iterations while gradually approaching the ground-truth distribution. Notably, this observation aligns with prior work~\cite{latentr3,rearec}.
\end{itemize}

\section{Related Work}
In this section, we navigate through existing research on LLM-based recommendation, and latent reasoning.

\vspace{+3pt}
\textbf{LLM-Based Recommendation} primarily utilize LLMs as the recommender backbone. By representing a user’s historical context in textual form and carefully designing prompts to guide the model’s behavior, these systems exploit the LLM’s broad world knowledge and reasoning capabilities to improve understanding of both item contents and user preferences~\cite{chatrec, d3, llara, tallrec, recllm-r1}.

Early work in LLM-based recommendation focused on either prompt engineering or direct fine-tuning with recommendation data. For instance, ChatRec~\cite{chatrec} positions the LLM as a decision-making engine: carefully designed prompts guide it to invoke traditional recommendation models to retrieve candidate items, which are then re-ranked and explained through additional prompts. In contrast, BIGRec~\cite{bigrec} directly fine-tunes the LLM on sequential recommendation tasks, using dot-product mapping to reduce hallucination. P5~\cite{p5} extends this approach by employing multi-task supervised learning—including sequential recommendation and content understanding—thereby substantially improving the LLM’s comprehension of recommendation scenarios.

Despite these achievements, direct training methods often constrain the model’s reasoning depth, limiting its ability to fully leverage test-time scaling potential. Motivated by the success of DeepSeek-R1~\cite{deepseek-r1}, recent studies have increasingly explored methods to stimulate LLM reasoning in recommendation tasks. For instance, R$^2$ec~\cite{r2rec} incorporates Chain-of-Thought (CoT) data into sequential recommendation, training the model with a language head to generate reasoning sequences and a recommendation head to predict target items. DMPO~\cite{dmpo} employs an optimized Direct Preference Optimization (DPO) strategy to enhance the LLM’s understanding of human preferences, resulting in improved reasoning and decision quality. Building on these ideas, RecLLM-R1~\cite{recllm-r1} introduces a two-stage training paradigm: an initial Supervised Fine-Tuning (SFT) phase, followed by Group Relative Policy Optimization (GRPO) with CoT-guided reasoning, aiming to jointly optimize recommendation accuracy, diversity, and other customized business objectives.

\vspace{+4pt}

\textbf{Latent Reasoning} is primarily characterized by decoupling the model’s thinking process from the explicit token embedding space~\cite{rearec, lares, latentr3, onepiece,flythinker}. By performing reasoning in a continuous latent space, the model’s upper-bound reasoning capacity can be better leveraged, while inference computational overhead is simultaneously reduced~\cite{simcot, coconut}. As an emerging research direction, several pioneering works have explored this paradigm in recommendation. For example, ReaRec~\cite{rearec} introduces multiple autoregressive reasoning steps prior to the decoding phase of SASRec~\cite{SASRec}, enabling the model to deliberate on user interests from multiple perspectives through consistent supervision with ground truth before producing a final prediction. Building on this idea, LARES~\cite{lares} extends the autoregressive process from single tokens to entire input sequences, employing dynamic reasoning lengths to better capture complex user interests. LatentR$^3$~\cite{latentr3} further employs a two-stage training paradigm of SFT followed by GRPO, where perplexity is used as a reward signal during GRPO to enhance the exploration capacity of latent reasoning in LLMs.
\section{Conclusion}

In this work, we introduced the concept of \emph{recommendation intuition} to guide latent reasoning in LLM-based recommendation. We presented \emph{IntuRec}, a two-stage framework that explicitly extracted and injected intuition to initialize the reasoning process from preference-aligned representations. By anchoring latent reasoning with a semantically grounded starting point, IntuRec mitigated the inefficiencies associated with unconstrained reasoning trajectories and enabled more effective exploration of user preferences. Overall, this study demonstrated that incorporating human-inspired intuition can substantially improve the alignment between latent reasoning processes and underlying user interests.

For future work, we plan to extend the evaluation of IntuRec to large-scale recommendation datasets to further explore its scalability and effectiveness. Additionally, we aim to investigate alternative methods for generating recommendation intuitions that are more efficient or semantically richer, potentially enhancing the quality and speed of latent reasoning.

\begin{acks}
This work was supported in part by the National Natural Science Foundation of China under Grant 62477001, in part by the State Key Laboratory of Complex \& Critical Software Environment under Grant SKLCCSE-2025ZX-12, and in part by the Engineering Research Center of Integration and Application of Digital Learning Technology, Ministry of Education under Grant 1441002.
\end{acks}

\bibliographystyle{ACM-Reference-Format}
\balance
\bibliography{8_ref}

\appendix
\section{More Dataset Validation}
\begin{table}[H]
\centering
\caption{Dataset statistics of Instruments and Books.}
\label{tab:data_statics_extra}
\resizebox{\columnwidth}{!}{%
\begin{tabular}{ccccccc}
\hline
Dataset     & \#Train  & \#Valid & \#Test  & \#User num & \#Item num & Sparsity \\ \hline
Instruments & 66500  & 8312  & 8313  & 12044    & 5030     & 99.84\%  \\
Books       & 682997 & 85375 & 85375 & 67708    & 41722    & 99.97\%  \\ \hline
\end{tabular}%
}
\end{table}
To further validate the effectiveness of IntuRec, we conduct experiments on two additional datasets, Instruments and Books. We compare IntuRec with BIGRec and LatentR$^3$. The statistical details of these datasets are summarized in Table \ref{tab:data_statics_extra}, and the experimental results are reported in Table \ref{tab:extra_dataset}.

\begin{table}[H]
\centering
\caption{Performance comparison between baselines and our proposed IntuRec on extra datasets. The best results are highlighted in bold.}
\label{tab:extra_dataset}
\resizebox{\columnwidth}{!}{%
\begin{tabular}{lcccc}
\toprule
Dataset & Metric & BIGRec & LatentR$^3$-B & IntuRec-B \\ 
\midrule
\multirow{2}{*}{Instruments} & Recall@5 & 0.0246 & 0.0251 & \textbf{0.0332} \\
 & NDCG@5 & 0.0187 & 0.0194 & \textbf{0.0263} \\ 
\midrule
\multirow{2}{*}{Books} & Recall@5 & 0.1032 & 0.1014 & \textbf{0.1041} \\
 & NDCG@5 & 0.0883 & 0.0880 & \textbf{0.0900} \\ 
\bottomrule
\end{tabular}%
}
\end{table}

As observed from the table, IntuRec consistently outperforms the baselines across these datasets, which further verifies the effectiveness of our proposed approach.

\section{Exploring Candidate List Generation via Traditional Recommendation Models}

To further explore the reliance of IntuRec on the candidate list for generating intuition embeddings, we attempt to use a traditional recommendation model (e.g., SASRec) for candidate generation. The corresponding variant is denoted as IntuRec-B+SASRec. The experimental results are shown in Table \ref{tab:sasrec_candidate}.

\begin{table}[H]
\centering
\caption{Performance comparison of candidate list generation using a traditional recommendation model (SASRec).}
\label{tab:sasrec_candidate}
\resizebox{\columnwidth}{!}{%
\begin{tabular}{lccc}
\toprule
Metric & BIGRec & IntuRec-B+SASRec & IntuRec-B \\ 
\midrule
Recall@5 & 0.0757 &0.0900 & \textbf{0.0986}	 \\
NDCG@5 & 0.0616 & 0.0736 & \textbf{0.0795} \\ 
\bottomrule
\end{tabular}
}
\end{table}

From the table, we can observe the following:
\begin{itemize}
    \item \textbf{IntuRec-B shows a significant advantage over IntuRec-B+SASRec.} This indicates that it is challenging to project the recommendation lists obtained from traditional models into the LLM recommender, which simultaneously verifies the rationality of the ISE design.
    \item \textbf{IntuRec-B+SASRec significantly outperforms BIGRec.} This suggests that the approach of using traditional models to provide candidate lists for IntuRec still holds potential for future improvements.
\end{itemize}

\section{Case Study}

To further illustrate the relationship among the recommendation list generated by IntuRec, the candidate list, and the ground truth, we extract a case study for better understanding. The results are presented in Table \ref{tab:case_study}.

\begin{table}[H]
\centering
\caption{Case study illustrating the generated candidate list and recommendation list relative to the user history and target item.}
\label{tab:case_study}

\resizebox{\columnwidth}{!}{%
\begin{tabular}{>{\centering\arraybackslash}m{0.25\linewidth} >{\centering\arraybackslash}m{0.68\linewidth}}
\toprule
\textbf{Category} & \textbf{Items} \\ 
\midrule

User History & 
\small
["Abbey Road", "Magical Mystery Tour", "Beatles '65", "Beatles For Sale"] \\ 
\midrule

Candidate List & 
\small
["Abbey Road", \textbf{"Beatles for Sale"}, "The Best of Four Tops: 20th Century Masters The Millennium Collection", "A Hard Day's Night Soundtrack The U.S. Album", "1967-1970 The Blue Album", "He ! Soundtrack The U.S. Album", "1962-1966", "Magical Mystery Tour", "Let It Bleed", "Hot In The Shade", "He !", "Please Please Me", "The Best Of The Moody Blues", \textbf{"The Beatles: Something New"}, \textbf{"Meet the Beatles"}] \\ 
\midrule

Target Item & 
\textbf{With the Beatles} \\ 
\midrule

Recommendation List& 
\small
["He !", \textbf{"With the Beatles"}, "Magical Mystery Tour", "Yellow Submarine Soundtrack", "Let It Be... Naked", "Abbey Road", "Beatles for Sale", "Meet the Beatles", "Hot Rocks", "The Best of Bread"] \\ 

\bottomrule
\end{tabular}%
}
\end{table}

As observed from the table, the candidate list shows a clear correlation with the ground truth, although it does not contain the exact target item. However, guided by the candidate list, IntuRec successfully hits the ground truth and ranks it highly.

\section{Efficiency and Cost}

To further evaluate the efficiency and cost of IntuRec, we profile its training and inference speed as well as memory consumption. The statistics are summarized in Table \ref{tab:efficiency}.

\begin{table}[H]
\centering
\caption{Efficiency and cost comparison among models during training and inference.}
\label{tab:efficiency}
\resizebox{\columnwidth}{!}{%
\begin{tabular}{lccc}
\toprule
\textbf{Metric} & \textbf{BIGRec} & \textbf{LatentR$^3$-B} & \textbf{IntuRec-B} \\ 
\midrule
\textbf{Train Ratio (s/it)} & 0.3559 & 0.5464 & 0.6191 \\
\textbf{Train Mem (MB)} & 25195 & 41373 & 33867 \\
\textbf{Inference Ratio (s/it)} & 1.2771 & 1.7520 & 1.6499 \\
\textbf{Test Mem (MB)} & 11223 & 47483 & 39979 \\ 
\bottomrule
\end{tabular}%
}
\end{table}

From the table, we can observe the following:
\begin{itemize}
    \item \textbf{Inference Advantage:} IntuRec-B is \textbf{faster} than LatentR$^3$-B. This is because LatentR$^3$-B must process an additional module (\textbf{LatentRTT}) at \textbf{every} autoregressive step, whereas IntuRec-B only performs the intuition aggregation at the first step.
    \item \textbf{Training Trade-off:} IntuRec-B has higher training latency because it utilizes the final-layer hidden states for backpropagation, which increases computational depth but is essential for the superior performance shown above.
\end{itemize}

Consequently, we believe this increased training cost is a well-justified investment, as it produces a model that is not only significantly more accurate but also more efficient during inference—a trade-off that is highly advantageous for latency-sensitive, real-world applications.

\end{document}